\documentclass[12pt]{iopart}

\usepackage{amssymb}
\usepackage{iopams}
\usepackage{graphicx}
\usepackage{epsfig}
\usepackage{epstopdf}
\begin{document}

\title{Multiple predator based capture process on complex networks}

\author{Rajput Ramiz Sharafat, Jie Li, Cunlai Pu, Rongbin Chen }

\address{School of Computer Science and Engineering, Nanjing University of Science and Technology, Nanjing 210094, China}
\ead{pucunlai@njust.edu.cn}
\vspace{10pt}
\begin{indented}
\item[]August 2016
\end{indented}

\begin{abstract}The predator/prey (capture) problem is a prototype of many network-related applications.  We study the capture process on complex networks by considering  multiple predators from multiple sources. In our model,  some lions start from  multiple sources simultaneously to capture the lamb by biased random walks, which are controlled with a free parameter $\alpha$. We derive  the  distribution of the lamb's lifetime and the expected lifetime $\left\langle T\right\rangle $. Through simulation, we find that the expected lifetime drops substantially with the increasing number of lions. We also study how the underlying topological structure affects the capture process, and obtain that locating on small-degree nodes is better than  large-degree nodes to prolong the lifetime of the lamb. Moreover, dense or homogeneous  network structures are against the survival of the lamb.
\end{abstract}

%
\section{Introduction}
Complex network modeling has been used in  many natural and artificial   systems, where  nodes represent individuals, and  edges  represent interactions between the individuals. Over the past decade, researchers have made great efforts to study the structural properties of complex networks \cite{barabasi2016network,Boccaletti20141,mendes2014structural}, the dynamical processes existing in complex networks \cite{barrat2008dynamical}, and how they affect each other \cite{PhysRevLett.111.128701,caldarelli2007large}. For instance, it was found that a variety of complex networks have similar statistical structural properties, such as small-world \cite{watts1998collective,muldoon2016small}, scale-free \cite{barabasi2009scale,PhysRevE.94.022302}, high clustering \cite{RevModPhys.74.47}, and so on.  Moreover, these general network properties were demonstrated to have great impacts on the network-based problems, such as traffic transport \cite{yan2006efficient,wang2006traffic,sole2016congestion,du2016physics,Pu2016261}, epidemic spreading \cite{RevModPhys.87.925,PhysRevE.93.032301,Pu2016129,Pu2015230,PhysRevE.91.062817}, network attacks \cite{PhysRevE.89.042811,pocock2012robustness,PhysRevE.66.065102,PhysRevE.65.056109,Pu20151633,Pu2015622,kovacs2015network}, synchronization \cite{Dörfler20141539,1440569} and control \cite{liu2011controllability,ruths2014control,Wang2002521,yan2015spectrum,yuan2013exact,Pu20124420}, link prediction \cite{Lü20111150,Cui2016202,Xu2016294}, etc. For example, one of the great findings is that the scale-free networks are robust to random attacks, but very fragile to target attacks \cite{albert2000error}. Another breakthrough is that the scale-free networks are proved to have very low ( even zero)  epidemic thresholds \cite{PhysRevLett.86.3200}.

Random walk is a basic process related to many of those network problems \cite{klafter2011first,white2015random}. In graph theory, random walk has been studied for decades, and thus the random walk characteristics like arrival time, meet time, commuting time, cover time, etc.,   \cite{lovasz1993random,0305-4470-38-8-R01}  have been well discussed.
In recent studies,  the random walk theory is used to explore the structures of complex networks such as network sampling \cite{PhysRevE.75.046114}, calculating node centrality \cite{Newman200539}, detecting cluster structures \cite{Rosvall29012008,PhysRevE.67.061901}, predicting missing links \cite{Backstrom:2011:SRW:1935826.1935914}, etc. Furthermore, random walk has widely applications in communication networks for  locating resources \cite{Millán2016165}, detecting replication attacks \cite{aalsalem2016new}, and designing navigation \cite{PhysRevE.80.016107,viswanathan1999optimizing} or routing protocols \cite{7426027}.
The predator-prey (lion-lamb) model is a common random walk model \cite{berryman1992orgins}. In the past, this model was used to illustrate the processes of energy transfer in  ecological systems \cite{:/content/aip/journal/jcp/76/7/10.1063/1.443410,barnes2010global}.  Recently, Sungmin et al \cite{PhysRevE.74.046118} studied the predator-prey model on complex networks, and they particularly focused on how the network structures affect the lamb survival probability.
Shao-Ping  et al \cite{Wang20084699} further utilized the predator-prey model in the search problems in point-to-point networks. However,  in both of their works, there is only one start point of predator.

In this paper, we generalize the predator-prey model by considering multiple start points of predators with biased random walks. We first derive the lifetime distribution as well as the expected lifetime of the lamb. Then, we study how the control parameter of the biased random walks and number of lions affect the expected lifetime of the lamb. Finally, we investigate how the underlying network structure  affects  the expected lifetime.

\section{Model}
We use the Price model \cite{price1976general} to generate the underlying scale-free networks. This model contains two steps:
(\romannumeral1) Growth: Initially, there are $m_0$ isolated nodes in the network. Then, each time we add a new node to the network with $m$ links going from the new node and pointing to the old nodes, where $m \leq m_0$.
(\romannumeral2)	Cumulative advantage: A new node points to an old node $i$ with probability $\Phi_i$:
\begin{equation}
\Phi_i=\frac{k_i^{in}+a}{\sum_j(k_j^{in}+a)},
\end{equation}
where $k_i^{in}$ is the indegree of node $i$, and $a$ is a given constant. Multiple connections from the new node to an old node are not allowed.  The process goes $N-m_0$ steps, where $N$ is desired network size.  We ignore the directions of links and obtain the undirected scale-free network. The average node degree is $\langle k \rangle=2m$, and the power-law parameter is $\gamma=2+a/m$.

In our capture model, initially some lions start from distinct source nodes to capture a lamb at the destination node. At each time step, every lion steps onto a neighbor node through biased random walk. For instance, a lion at node $u$  steps onto a neighbor node $i$ of $u$ with the probability:
\begin{equation}
\phi_{ui}=\frac{k_i^{\alpha}}{\sum_jk_j^{\alpha}},
\end{equation}
Where $k_j$ is the degree of $j$, and the sum runs over all the neighbor nodes of $u$. $\alpha$ is a control parameter, when $\alpha>0$, the random walk is biased to large-degree nodes, while when $\alpha<0$, the random walk is biased to small-degree nodes. When $\alpha=0$, the biased random walk is degenerated to simple random walk.
The capture process will end when the lamb is captured by the lion which first arrives at the destination node.
\section{Analytical results}

Assuming a graph $G(N,M)$, where $N$ and $M$ represent the sets of nodes and edges respectively. The adjacency relationship of $G$ can be represented by a Matrix $\textbf{A}$. If there is a link between node $i$ and node $j$, then $A_{ij}=k_j^{\alpha}$. Here $A_{ij}$ can be taken as the directed weight of the link in that $A_{ij}$ is generally not equal to $ A_{ji}$. When $\alpha=0$, $A_{ij}=A_{ji}=1$, which is the same as the  general  adjacency matrix. If there is no link between $i$ and $j$, $A_{ij}=A_{ji}=0$, which is also the same as the general adjacency matrix. Then, the strength of node $i$ is $s_i=\sum A_{ij}$. Furthermore, the transition probability matrix is $\textbf{P}=\textbf{K}^{-1}\textbf{A}$, where $\textbf{K}=diag(s_1,s_2,\dots,s_N)$.   Assuming that at time $t=0$, a lion  on node $u$ begins to perform the biased random walk, then the probability of reaching node  $v$  after  $t$  steps of walks can be described as  below:

 \begin{equation}
p_{uv}(t)=(\textbf{P}^t)_{uv}=\sum_{j_{1}..j_{t-1}}\frac{A_{uj_{1}}}{s_{u}}\cdot \frac{A_{j_{1}j_{2}}}{s_{j_{1}}}\cdot\cdot\cdot \frac{A_{j_{t-1}v}}{s_{j_{t-1}}}=\phi_{uj_1}\phi_{j_1j_2}\cdots\phi_{j_{t-1}v},
\end{equation}
where $j_i$ is the intermediate node  the lion visits at time $i$. Note that the lion may visit $v$ many times in $t$ steps. According to our model, the capture process ends when the destination node (the location of the lamb) is first visited by the lion. To compute the first arrival time, we set the $v$th column of $\textbf{P}$ to be zero, and get the new transition probability matrix as:
\begin{equation}
\textbf{Q}(v)=(\textbf{p}_1, \dots, \textbf{p}_{v-1}, 0, \textbf{p}_{v+1},\dots, \textbf{p}_N).
\end{equation}
 We can infer  through the new transition probability matrix that  the probability  the lion leaves $v$ after first arriving is zero.  Then, we can compute the probability that $v$ has not been visited in the $t$ steps, or in other words, the probability that the lifetime of the lamb is larger than $t$, which is as follows:
 \begin{equation}
Prob\{T_{uv}>t\})=\sum_{j=1}^{N}(\textbf{Q}(v)^{t})_{uj},  \:t=0,1,2,....
\end{equation}
Next, we assume that initially there are many lions starting from multiple source nodes, which are marked as  $1, 2, \dots, z$ ($z<N-1$), and the lamb is still on node $v$. Note that $v$ is not among the $z$ source nodes. The numbers of lions on the $z$ source nodes are $n_1, n_2,\dots, n_z$ respectively. We define $T_{iv}^j$ as the time when the $j$th lion  on source node $i$ first arrives at node $v$. Then, the lifetime of the lamb $T_v$ is the smallest first arrival time of all the lions, which can be described as follows:
\begin{equation}
T_v=\min\{T_{1v}^1, T_{1v}^2,\dots, T_{1v}^{n_1}, \dots, T_{zv}^1, T_{zv}^2, \dots, T_{zv}^{n_z}\}.
\end{equation}
The probability that the lifetime of the lamb is larger than $t$ is:
\begin{eqnarray}
Prob\{T_{v}>t\}&=Prob\{\min\{T_{1v}^{1},T_{1v}^{2},\dots,T_{1v}^{n_{1}},\dots,T_{zv}^{1},T_{zv}^{2},\dots,T_{zv}^{n_{z}}\}>t\} \nonumber\\
&=(\prod_{i}^{n_{1}}prob\{T_{1v}^{i}>t\})\cdots(\prod_{i}^{n_{z}}prob\{T_{zv}^{i}>t\}).
\end{eqnarray}
Since each lion performs independent biased random walks, we have:
\begin{equation}
Prob\{T_{v}>t\}=\prod_{i={1}}^{z}(prob\{T_{iv}>t\})^{n_{i}}.
\end{equation}
Combining Eq. 5 and Eq. 8, we have:
\begin{equation}
Prob\{T_{v}>t\}=\prod_{i={1}}^{z}(\sum_{j=1}^{N}(\textbf{Q}(v)^{t}) _{ij})^{n_{i}}, \:t=0,1,2,....
\end{equation}
Then, the distribution of the lamb's lifetime  can be calculated as follows:
\begin{eqnarray}
Prob\{T_{v}=t\}= Prob\{T_{v}>t-1\}-Prob\{T_{v}>t\}\nonumber \\
=\prod_{i={1}}^{z}(\sum_{j=1}^{N}(\textbf{Q}(v)^{t-1})_{ij})^{n_{i}}-\prod_{i={1}}^{z}(\sum_{j=1}^{N}(\textbf{Q}(v)^{t})_{ij})^{n_{i}},  \:t=0,1,2,....
\end{eqnarray}
Finally, the expected lifetime of the lamb is calculated as below:
\begin{eqnarray}
\langle T_{v} \rangle &= \sum_{t=0}^{\infty}t\cdot Prob\{T_{v}=t\}\nonumber \\
&=\sum_{t=0}^{\infty}Prob\{T_{v}>t\}\nonumber\\
&=\sum_{t=0}^{\infty}\prod_{i={1}}^{z}(\sum_{j=1}^{N}(\textbf{Q}(v)^{t})_{ij})^{n_{i}}.
\end{eqnarray}

The above derivation provides a general method for calculating the expected lifetime for all types of graphs. However,  we have more simpler calculation methods for special graphs.
\subsection{ Fully connected graph}
Let's consider a fully connected graph, in which each node is connected to every other node with an edge, and  thus the degrees of all nodes are identical and equal $N-1$. Assuming there are $n$ lions and a lamb randomly distributed in the fully connected graph. Note that, initially, the lions'   sites may be overlap or not, and the only constraint is that the lamb's site is not occupied by any lion at the beginning. For this case, we have $Prob\{T>0\}=1$, $Prob\{T>1\}=(1-\frac{1}{N-1})^n$, and $Prob\{T>t\}=(1-\frac{1}{N-1})^{tn}$. Furthermore, we calculate the expected lifetime as follows:
\begin{eqnarray}
\langle T \rangle=\sum_{t=1}^{\infty}t\left[Prob(T>t-1)-Prob(T>t)\right] \\
= \sum_{t=1}^{\infty}t\left[\left( 1-\frac{1}{N-1}\right)^{(t-1)n}-\left(1-\frac{1}{N-1}\right)^{tn}\right]\\
\!=\! \left[1\!-\!\left(1\!-\!\frac{1}{N\!-\!1}\right)^{n}\right]
\!+\!\sum_{t=1}^{\infty}(t\!+\!1)\left[ \left(1\!-\!\frac{1}{N\!-\!1}\right)^{tn}\!-\!\left(1\!-\!\frac{1}{N\!-\!1}\right)^{(t\!+\!1)n}\right]
\end{eqnarray}
Based on Eq. 13, we have:
\begin{eqnarray}
\left(1-\frac{1}{N-1}\right)^n\langle T \rangle=\sum_{t=1}^{\infty}t\left[\left(1-\frac{1}{N-1}\right)^{tn}-\left(1-\frac{1}{N-1}\right)^{(t+1)n}\right].
\end{eqnarray}
Eq. 14 minus Eq. 15 yields:
\begin{eqnarray}
\left[1-\left(1-\frac{1}{N-1}\right)^n\right]\langle T \rangle
=\left[1-\left(1-\frac{1}{N-1}\right)^n\right] \nonumber\\
+\sum_{t=1}^{\infty}\left[\left(1-\frac{1}{N-1}\right)^{tn}-\left(1-\frac{1}{N-1}\right)^{(t+1)n}\right]\nonumber\\
=\left[1\!-\!\left(1\!-\!\frac{1}{N\!-\!1}\right)^n\right]+ \sum_{t=1}^{\infty}\left(1\!-\!\frac{1}{N\!-\!1}\right)^{tn}\!-\!\sum_{t=2}^{\infty}\left(1-\frac{1}{N-1}\right)^{tn}\nonumber\\
=\left[1-\left(1-\frac{1}{N-1}\right)^n\right]+\frac{\left(1-\frac{1}{N-1}\right)^{n}}{1-\left(1-\frac{1}{N-1}\right)^{n}}-\frac{\left(1-\frac{1}{N-1}\right)^{2n}}{1-\left(1-\frac{1}{N-1}\right)^{n}}\nonumber\\
=\left[1-\left(1-\frac{1}{N-1}\right)^n\right]+\left(1-\frac{1}{N-1}\right)^{n}=1.
\end{eqnarray}
Then, we obtain:
\begin{eqnarray}
\langle T \rangle=\frac{1}{1-\left(1-\frac{1}{N-1}\right)^n}.
\end{eqnarray}
According to Eq. 17, when $n=1$, $\langle T \rangle=N-1$, while when $n$ is very large, we obtain:
\begin{eqnarray}
\langle T \rangle\approx\frac{1}{1-\left(1-\frac{n}{N-1}\right)}=\frac{N-1}{n}.
\end{eqnarray}
From Eq. 18, we see that the expected lifetime is approximately proportional to the network size and  inversely proportional to the number of lions.
\subsection{ ER random graph}

For the ER random graph \cite{erdos1960evolution}, the degree distribution obeys the Possion distribution. We assume that the degree of all nodes are  $\langle k \rangle$. We know that for the ER random graph, the probability for any two nodes being connected with an edge is:  
\begin{equation}
p \approx \frac{N\langle k \rangle/2}{N(N-1)/2}=\frac{\langle k \rangle}{N-1}.
\end{equation}
Then, we have $Prob\{T>0\}=1$, and 
$Prob\{T>1\}=\left(1-\frac{\langle k \rangle}{N-1}\frac{1}{\langle k \rangle }\right)^n=\left(1-\frac{1}{N-1}\right)^n$, and $Prob\{T>t\}=\left(1-\frac{1}{N-1}\right)^{tn}$, which are all the same as the fully connected graph. Thus, Eq. 17 and 18 are also applicable to the ER random graph when $\langle k \rangle$ is large. When $\langle k \rangle$ is   small,  approximating $1/k_i$ (the degree of node $i$) with $1/\langle k \rangle$ is not suitable in the calculation. Thus, Eq. 17 and 18 are not applicable to the ER random graph, when $\langle k \rangle$ is small. 
\section{Simulation results}
In this section, we mainly study how the lamb's expected lifetime affected by the factors in our model such as the control parameter of the biased random walks, the number of lions, as well as the degree of the node, where the lamb locates on. In addition, we investigate how the average node degree and the degree distribution of the underlying networks influence the expected lifetime.

First, we study the control parameter $\alpha$ of the biased random walks, and at the same time testify the analytical results by simulation. We use the Price model to generate the underlying scale-free network  which contains 100 nodes with a average degree of 4 and a power-law parameter of 2.5, shown in Fig. 1. Then, we set node 40 as the destination node where the lamb locates on,  node 10 and node 90 as the lions' starting points, on which the numbers of lions are 1 and 2 respectively. We let the lions perform the biased random walks to capture the lamb, and record the lamb's lifetime. We perform the simulation $10^4$ times and calculate the average (expected) lifetime $\langle T \rangle$. The analytical and simulation results are given in Fig. 1. We see that $\langle T \rangle$ decreases first and then increases with $\alpha$, and there is an optimal $\alpha$ corresponding to the minimum $\langle T \rangle$. Moreover, the analytical results and the simulation results agree very well with each other.
\begin{figure}
\centering
\includegraphics[width=0.8\textwidth]{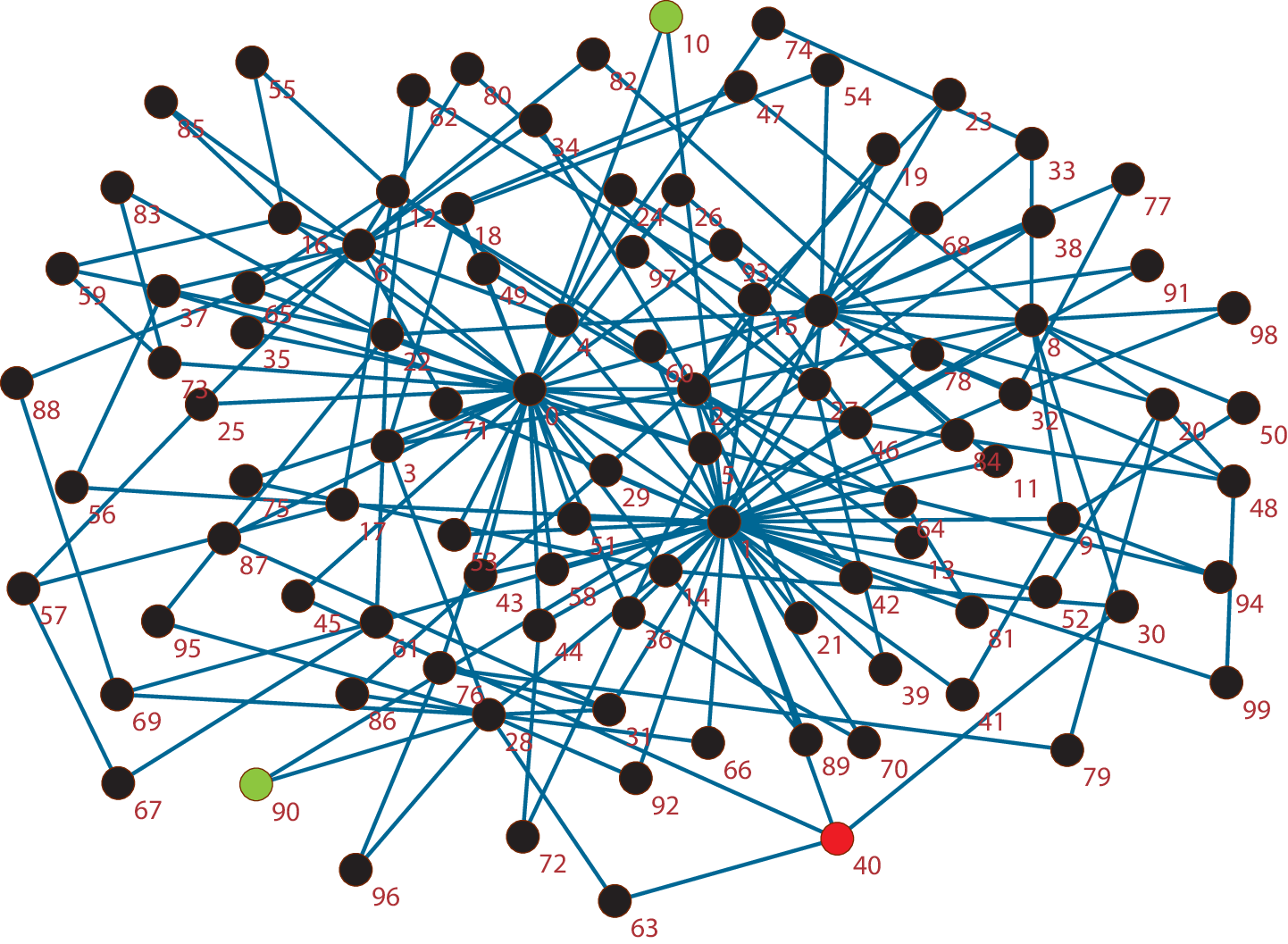}
\caption{A small network generated by the Price model. $N=100$, $\langle k \rangle=4$, and $\gamma=2.5$. The lamb's site is node 40, and the lions sites are node 10 and 90 with one lion on each site. }

%
\label{fig:1eps}       
\end{figure}
\begin{figure}
	\centering
	\includegraphics[width=0.8\textwidth]{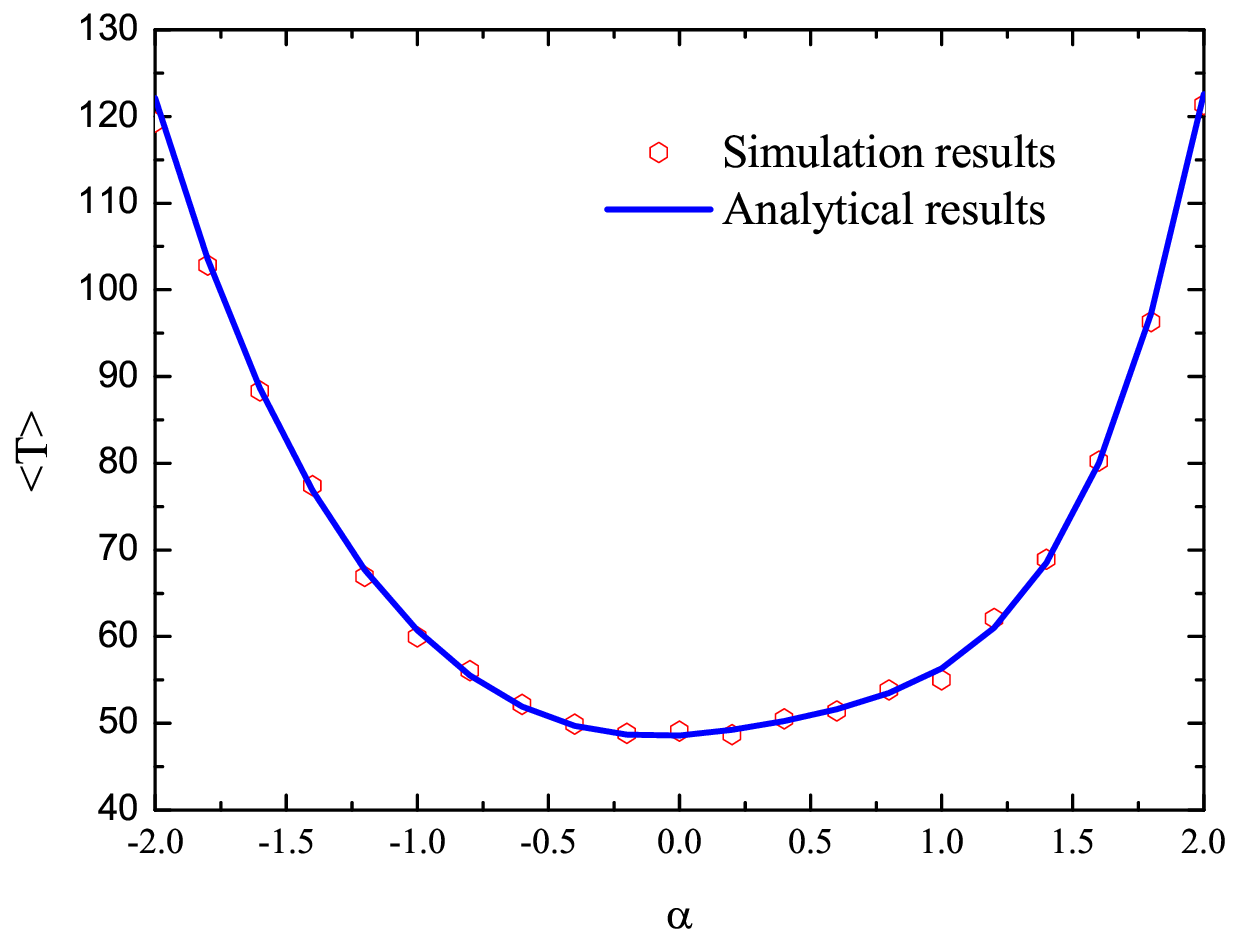}
	\caption{The lamb's expected lifetime $\langle T \rangle$ vs. the parameter   $\alpha$ of the biased random walks. The underlying network is in Fig. 1.  Each data point is the average of $10^4$ independent runs. }
	
	%
	\label{fig:1eps}       
\end{figure}

Then, we investigate how the number of lions affects the expected lifetime. In the simulation, we randomly select a node as the lamb's site and another $n$ lions' sites with one lion on each site. Under this condition, we perform the capture process and calculate the expected lifetime, which is given in Fig. 3(a). Clearly, we see that $\langle T \rangle$ decreases with $n$ abruptly and then converges. Through the above derivation of the lamb's expected lifetime, we know that the lifetime of the lamb equals the first arrival time of the luckiest lion  among all the lions who finds the lamb first. Generally, the more lions, the smaller the lamb's lifetime, which agrees with the analytical results of the fully connected graph and the ER random graph (Eq. 18).   Next, we discuss the influence of the degree of the lamb's site, denoted by $k_{\textnormal{lamb}}$. In the simulation, we select the lamb's site based on its degree, and then randomly select another 10 sites with one lion on each site. The control parameter $\alpha$ is set to be zero. In this case, we calculate the expected lifetime as a function of the degree of the lamb's site, which is shown in Fig. 3(b). Obviously, we see that $\langle T \rangle$ decreases with increasing $ k_{\textnormal{lamb}}$, which means that locating on large-degree sites is adverse to the survival of the lamb. In other words, large-degree nodes are easier to be visited by random walkers than small-degree nodes, which is also obtained in Ref.\cite{PhysRevLett.92.118701}.
\begin{figure}[h!]
	\centering
	\includegraphics[width=0.8\textwidth]{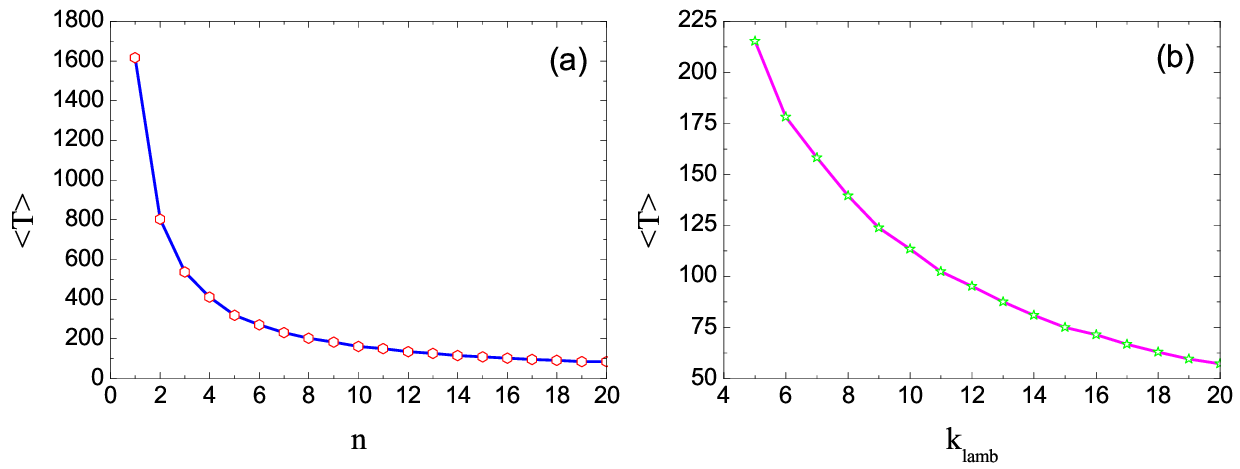}
	\caption{The lamb's expected lifetime $\langle T \rangle$ vs. (a) the number of lions $n$ and (b) the degree of the lamb's site $k_{\textnormal{lamb}}$. The underlying network is generated by the Price model. The parameters  are $N=1000$, $\langle k \rangle=10$, $\gamma=2.5$,  $\alpha=0$ and $n=10$ for (b).  The details about the sites selection for the lamb and lions are illustrated in the text.   Each data point is the average of $10^4$ independent runs. }
	
	%
	\label{fig:1eps}       
\end{figure} 

Finally, we discuss how the network structure affects the lifetime of the lamb. We focus on the average node degree and the degree distribution, and study one network parameter by fixing the others. In the simulation, we randomly select a node as the lamb's site, and another 10 nodes as the lions' sites with one lion on each site. The biased random walk parameter $\alpha$ is set be zero. Simulation results are given in Fig. 4, in which each data point is the average of $10^4$ independent runs. We see the similar trend of the curves of $\langle T \rangle$, which go down fastly and then tend to be stable with increasing $\langle k \rangle$ in Fig. 4(a) and $\gamma$ in Fig. 4(b).  When the network becomes more denser, the network diameter becomes smaller, which means the  lions and the lamb are more closer to each other, and this leads to the decrease of the lamb's lifetime.  When the network are very dense, $\langle k \rangle $ does not have significant influence on $\langle T \rangle$, which can also be predicted by Eq. 18. Furthermore, from Fig. 3(b) we know that random walkers are prone to visit large-degree nodes.  If the lamb locates on one of the small-degree nodes, which constitute the largest part of  scale-free networks,  it is relatively hard for the lions to catch the lamb. Thus, when the network becomes more homogeneous (by increasing $\gamma$), the random walkers will visit  all nodes more fairly, which increases the capture efficiency or decreases the lamb's lifetime.
\begin{figure}[h!]
	\centering
	\includegraphics[width=0.8\textwidth]{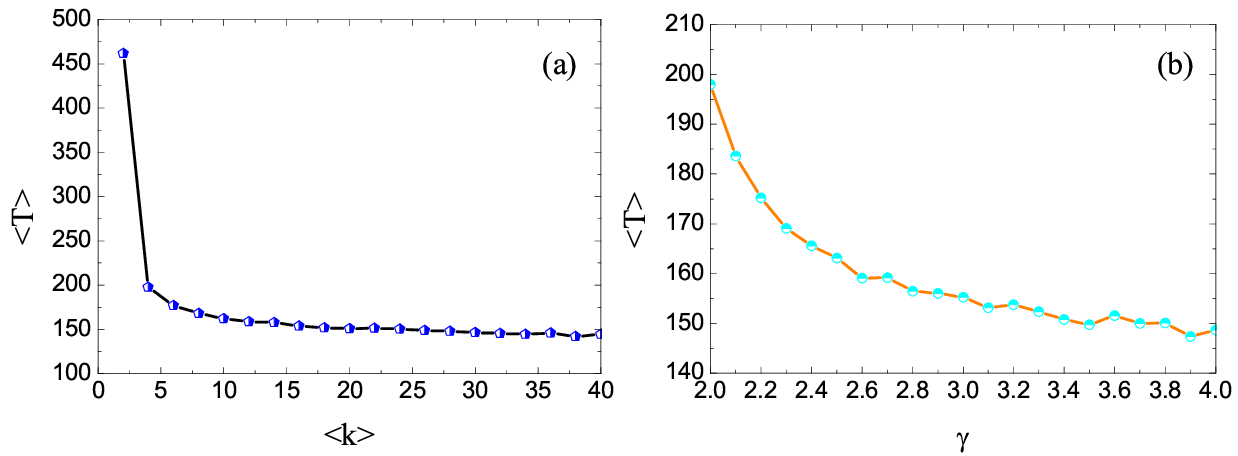}
	\caption{The lamb's expected lifetime $\langle T \rangle$ vs. (a) the average node degree $\langle k \rangle$ and (b) the power-law parameter $\gamma$. The underlying networks are  generated by the Price model. The parameters  are $N=1000$, $\gamma=2.5$ for (a), $\langle k \rangle=10$ for (b),  $\alpha=0$ and $n=10$.  The details about the sites selection for the lamb and lions are illustrated in the text.   Each data point is the average of $10^4$ independent runs. }
	
	%
	\label{fig:1eps}       
\end{figure}
\section{Conclusion}
In summary, we study a new kind of capture process, in which there are many lions starting from different source nodes. In detail, we derive the distribution of the lamb's lifetime as well as the expected lifetime. For the fully connected graph and ER random graph, we provide a simple approximate calculation of the expected lifetime, and find that the expected lifetime is proportional to the network size and inversely proportional to the number of lions. Next,   we study how the factors in our model affect the capture process on  scale-free networks  by simulation. We find that different values of the biased random walks parameter lead to different capture efficiency. Generally, given the source and destination nodes, there is always optimal control parameter corresponding to the smallest lifetime, or the highest capture efficiency. Furthermore, we obtain that the more lions, the faster to capture the lamb. Especially, when the number of lions is small, a small addition of lions results in larger improvement of capture efficiency. We also find that the larger the degree of the lamb's site, the smaller the expected lifetime. Finally, we study how the network structure affects the capture process by simulation. We find that the more denser or the more homogeneous the underlying network is, the smaller the lifetime of the lamb is. We believe that our model is not only interesting in itself, but also provides some clues to the designing of search strategies in P2P networks.
\newline
\section*{Acknowledgments}
This work was  supported by the National Natural Science Foundation of China (Grant No.  61304154), the Specialized Research Fund for the Doctoral Program of Higher Education of China  (Grant No. 20133219120032),  the Postdoctoral Science Foundation of China (Grant No. 2013M541673), and China Postdoctoral Science Special Foundation (Grant No. 2015T80556).
\newline


\begin{thebibliography}{10}
\providecommand{\url}[1]{\texttt{#1}}
\providecommand{\urlprefix}{URL }
\expandafter\ifx\csname urlstyle\endcsname\relax
  \providecommand{\doi}[1]{doi:\discretionary{}{}{}#1}\else
  \providecommand{\doi}{doi:\discretionary{}{}{}\begingroup
  \urlstyle{rm}\Url}\fi
\providecommand{\eprint}[2][]{\url{#2}}
\bibitem{barabasi2016network}
Barab\'{a}si A L. Network science[M]. Cambridge University Press, 2016.

\bibitem{Boccaletti20141}
Boccaletti S, Bianconi G, Criado R, et al. The structure and dynamics of multilayer networks[J]. Physics Reports, 2014, 544(1): 1-122.

\bibitem{mendes2014structural}
Mendes J F F, Dorogovtsev S N, Goltsev A V. Structural properties of complex networks[J]. SUMMERSOLSTICE 2014, 2014: 27.

\bibitem{barrat2008dynamical}
Barrat A, Barthelemy M, Vespignani A. Dynamical processes on complex networks[M]. Cambridge University Press, 2008.

\bibitem{PhysRevLett.111.128701}
Granell C, G\'{o}mez S, Arenas A. Dynamical interplay between awareness and epidemic spreading in multiplex networks[J]. Physical review letters, 2013, 111(12): 128701.

\bibitem{caldarelli2007large}
Caldarelli G, Vespignani A. Large scale structure and dynamics of complex networks[M]. World Scientific, 2007

\bibitem{watts1998collective}
Watts D J, Strogatz S H. Collective dynamics of 'small-world'networks[J]. nature, 1998, 393(6684): 440-442.

\bibitem{muldoon2016small}
Muldoon S F, Bridgeford E W, Bassett D S. Small-World Propensity and Weighted Brain Networks[J]. Scientific reports, 2016, 6.

\bibitem{barabasi2009scale}
Barab\'{a}si A L. Scale-free networks: a decade and beyond[J]. science, 2009, 325(5939): 412-413.

\bibitem{PhysRevE.94.022302}
Tim\'{a}r G, Dorogovtsev S N, Mendes J F F. Scale-free networks with exponent one[J]. Physical Review E, 2016, 94(2): 022302.

\bibitem{RevModPhys.74.47}
Albert R, Barab\'{a}si A L. Statistical mechanics of complex networks[J]. Reviews of modern physics, 2002, 74(1): 47.

\bibitem{yan2006efficient}
Yan G, Zhou T, Hu B, et al. Efficient routing on complex networks[J]. Physical Review E, 2006, 73(4): 046108.

\bibitem{wang2006traffic}
Wang W X, Wang B H, Yin C Y, et al. Traffic dynamics based on local routing protocol on a scale-free network[J]. Physical Review E, 2006, 73(2): 026111.

\bibitem{sole2016congestion}
Sol\'{e}-Ribalta A, G\'{o}mez S, Arenas A. Congestion induced by the structure of multiplex networks[J]. Physical review letters, 2016, 116(10): 108701.

\bibitem{du2016physics}
Du W B, Zhou X L, Jusup M, et al. Physics of transportation: Towards optimal capacity using the multilayer network framework[J]. Scientific reports, 2016, 6.

\bibitem{Pu2016261}
Pu C, Li S, Yang X, et al. Information transport in multiplex networks[J]. Physica A: Statistical Mechanics and its Applications, 2016, 447: 261-269.

\bibitem{RevModPhys.87.925}
Pastor-Satorras R, Castellano C, Van Mieghem P, et al. Epidemic processes in complex networks[J]. Reviews of modern physics, 2015, 87(3): 925.

\bibitem{PhysRevE.93.032301}
Shen Z, Cao S, Wang W X, et al. Locating the source of diffusion in complex networks by time-reversal backward spreading[J]. Physical Review E, 2016, 93(3): 032301.

\bibitem{Pu2016129}
Pu C, Li S, Yang X X, et al. Traffic-driven SIR epidemic spreading in networks[J]. Physica A: Statistical Mechanics and its Applications, 2016, 446: 129-137.

\bibitem{Pu2015230}
Pu C, Li S, Yang J. Epidemic spreading driven by biased random walks[J]. Physica A: Statistical Mechanics and its Applications, 2015, 432: 230-239.

\bibitem{PhysRevE.91.062817}
Yang H X, Tang M, Lai Y C. Traffic-driven epidemic spreading in correlated networks[J]. Physical Review E, 2015, 91(6): 062817.

\bibitem{PhysRevE.89.042811}
Min B, Do Yi S, Lee K M, et al. Network robustness of multiplex networks with interlayer degree correlations[J]. Physical Review E, 2014, 89(4): 042811.

\bibitem{pocock2012robustness}
Pocock M J O, Evans D M, Memmott J. The robustness and restoration of a network of ecological networks[J]. Science, 2012, 335(6071): 973-977.

\bibitem{PhysRevE.66.065102}
 Motter A E, Lai Y C. Cascade-based attacks on complex networks[J]. Physical Review E, 2002, 66(6): 065102.

\bibitem{PhysRevE.65.056109}
Holme P, Kim B J, Yoon C N, et al. Attack vulnerability of complex networks[J]. Physical Review E, 2002, 65(5): 056109.

\bibitem{Pu20151633}
Pu C, Li S, Michaelson A, et al. Iterative path attacks on networks[J]. Physics Letters A, 2015, 379(28): 1633-1638.

\bibitem{Pu2015622}
Pu C L, Cui W. Vulnerability of complex networks under path-based attacks[J]. Physica A: Statistical Mechanics and its Applications, 2015, 419: 622-629.

\bibitem{kovacs2015network}
Kov\'{a}cs I A, Barab\'{a}si A L. Network science: Destruction perfected[J]. Nature, 2015, 524(7563): 38-39.

\bibitem{Dörfler20141539}
D\"{o}rfler F, Bullo F. Synchronization in complex networks of phase oscillators: A survey[J]. Automatica, 2014, 50(6): 1539-1564.

\bibitem{1440569}
Lu J, Chen G. A time-varying complex dynamical network model and its controlled synchronization criteria[J]. IEEE Transactions on Automatic Control, 2005, 50(6): 841-846.

\bibitem{liu2011controllability}
Liu Y Y, Slotine J J, Barabási A L. Controllability of complex networks[J]. Nature, 2011, 473(7346): 167-173.

\bibitem{ruths2014control}
Ruths J, Ruths D. Control profiles of complex networks[J]. Science, 2014, 343(6177): 1373-1376.

\bibitem{Wang2002521}
Wang X F, Chen G. Pinning control of scale-free dynamical networks[J]. Physica A: Statistical Mechanics and its Applications, 2002, 310(3): 521-531.

\bibitem{yan2015spectrum}
Yan G, Tsekenis G, Barzel B, et al. Spectrum of controlling and observing complex networks[J]. Nature Physics, 2015, 11(9): 779-786.

\bibitem{yuan2013exact}
Yuan Z, Zhao C, Di Z, et al. Exact controllability of complex networks[J]. Nature communications, 2013, 4.

\bibitem{Pu20124420}
Pu C L, Pei W J, Michaelson A. Robustness analysis of network controllability[J]. Physica A: Statistical Mechanics and its Applications, 2012, 391(18): 4420-4425.

\bibitem{Lü20111150}
L\"{u} L, Zhou T. Link prediction in complex networks: A survey[J]. Physica A: Statistical Mechanics and its Applications, 2011, 390(6): 1150-1170.

\bibitem{Cui2016202}
Cui W, Pu C, Xu Z, et al. Bounded link prediction in very large networks[J]. Physica A: Statistical Mechanics and its Applications, 2016, 457: 202-214.

\bibitem{Xu2016294}
Xu Z, Pu C, Yang J. Link prediction based on path entropy[J]. Physica A: Statistical Mechanics and its Applications, 2016, 456: 294-301.

\bibitem{albert2000error}
Albert R, Jeong H, Barab\'{a}si A L. Error and attack tolerance of complex networks[J]. nature, 2000, 406(6794): 378-382.

\bibitem{PhysRevLett.86.3200}
Pastor-Satorras R, Vespignani A. Epidemic spreading in scale-free networks[J]. Physical review letters, 2001, 86(14): 3200.

\bibitem{klafter2011first}
Klafter J, Sokolov I M. First steps in random walks: from tools to applications[M]. Oxford University Press, 2011.

\bibitem{white2015random}
White R T. Random Walks on Random Lattices and Their Applications[D]. Florida Institute of Technology, 2015.

\bibitem{lovasz1993random}
Lov\'{a}sz L. Random walks on graphs[J]. Combinatorics, Paul erdos is eighty, 1993, 2: 1-46.

\bibitem{0305-4470-38-8-R01}
Burioni R, Cassi D. Random walks on graphs: ideas, techniques and results[J]. Journal of Physics A: Mathematical and General, 2005, 38(8): R45.

\bibitem{PhysRevE.75.046114}
Yoon S, Lee S, Yook S H, et al. Statistical properties of sampled networks by random walks[J]. Physical Review E, 2007, 75(4): 046114.

\bibitem{Newman200539}
Newman M E J. A measure of betweenness centrality based on random walks[J]. Social networks, 2005, 27(1): 39-54.

\bibitem{Rosvall29012008}
Rosvall M, Bergstrom C T. Maps of random walks on complex networks reveal community structure[J]. Proceedings of the National Academy of Sciences, 2008, 105(4): 1118-1123.


\bibitem{PhysRevE.67.061901}
Zhou H. Distance, dissimilarity index, and network community structure[J]. Physical review e, 2003, 67(6): 061901.

\bibitem{Backstrom:2011:SRW:1935826.1935914}
Backstrom L, Leskovec J. Supervised random walks: predicting and recommending links in social networks[C]//Proceedings of the fourth ACM international conference on Web search and data mining. ACM, 2011: 635-644.

\bibitem{Millán2016165}
Mill\'{a}n V M L, Cholvi V, Anta A F, et al. Resource location based on precomputed partial random walks in dynamic networks[J]. Computer Networks, 2016, 103: 165-180.

\bibitem{aalsalem2016new}
Aalsalem M Y, Khan W Z, Saad N M, et al. A New Random Walk for Replica Detection in WSNs[J]. PloS one, 2016, 11(7): e0158072.

\bibitem{PhysRevE.80.016107}
Fronczak A, Fronczak P. Biased random walks in complex networks: The role of local navigation rules[J]. Physical Review E, 2009, 80(1): 016107.

\bibitem{viswanathan1999optimizing}
Viswanathan G M, Buldyrev S V, Havlin S, et al. Optimizing the success of random searches[J]. Nature, 1999, 401(6756): 911-914.

\bibitem{7426027}
Huang B, Feng Y, Li X, et al. An angle-based directed random walk privacy enhanced routing protocol for WSNs[C]//2015 International Conference on Information and Communications Technologies (ICT 2015). IET, 2015: 1-5.

\bibitem{berryman1992orgins}
Berryman A A. The Orgins and Evolution of Predator-Prey Theory[J]. Ecology, 1992, 73(5): 1530-1535.

\bibitem{:/content/aip/journal/jcp/76/7/10.1063/1.443410}
Zumofen G, Blumen A. Energy transfer as a random walk. II. Two-dimensional regular lattices[J]. The Journal of Chemical Physics, 1982, 76(7): 3713-3731.

\bibitem{barnes2010global}
Barnes C, Maxwell D, Reuman D C, et al. Global patterns in predator-prey size relationships reveal size dependency of trophic transfer efficiency[J]. Ecology, 2010, 91(1): 222-232.

\bibitem{PhysRevE.74.046118}
Lee S, Yook S H, Kim Y. Diffusive capture process on complex networks[J]. Physical Review E, 2006, 74(4): 046118.
		
\bibitem{Wang20084699}
Wang S P, Pei W J. First passage time of multiple Brownian particles on networks with applications[J]. Physica A: Statistical Mechanics and its Applications, 2008, 387(18): 4699-4708.

\bibitem{price1976general}
Derek de Solla Price. A general theory of bibliometric and other cumulative advantage processes. Journal of the American Society for Information Science, 27:292-306, 1976.

\bibitem{erdos1960evolution}
Erd\"{o}s P, R\'{e}nyi A. On the evolution of random graphs[J]. Publ. Math. Inst. Hung. Acad. Sci, 1960, 5(17-61): 43.

		
\bibitem{PhysRevLett.92.118701}
Noh J D, Rieger H. Random walks on complex networks[J]. Physical review letters, 2004, 92(11): 118701.


\end{thebibliography}
\end{document}